# Calculating the Middle Ages?
# The Project »Complexities and Networks in the Medieval Mediterranean and the Near East« (COMMED)

## Johannes Preiser-Kapeller*

The project »Complexities and networks in the Medieval Mediterranean and Near East« (*COMMED*) at the Division for Byzantine Research of the Institute for Medieval Research (*IMAFO*) of the Austrian Academy of Sciences focuses on the adaptation and development of concepts and tools of network theory and complexity sciences for the analysis of societies, polities and regions in the medieval world in a comparative perspective. Key elements of its methodological and technological toolkit are applied, for instance, in the new project »Mapping medieval conflicts: a digital approach towards political dynamics in the pre-modern period« (*MEDCON*), which analyses political networks and conflict among power elites across medieval Europe with five case studies from the 12th to 15th century. For one of these case studies on 14th century Byzantium, the explanatory value of this approach is presented in greater detail. The presented results are integrated in a wider comparison of five late medieval polities across Afro-Eurasia (Byzantium, China, England, Hungary and Mamluk Egypt) against the background of the »Late Medieval Crisis« and its political and environmental turmoil. Finally, further perspectives of *COMMED* are outlined.

*Keywords: complexity theory; network analysis; quantitative methods; comparative history; global history; Byzantine history; Mediterranean studies; environmental studies; climate history; social theory.*

## Introduction

In the year 2012, »Complexities and networks in the Medieval Mediterranean and Near East« (*COMMED*) was established as a research project at the Division for Byzantine Research of the Institute for Medieval Research (*IMAFO*) of the Austrian Academy of Sciences. Its aim is the integration, adaptation and further development of concepts and tools of network theory and complexity sciences for the analysis of the Byzantine Empire and neighbouring polities and regions in the medieval Mediterranean and Near East. With the help of these instruments, social, economic, religious, political and intellectual entanglements between individuals, groups, communities, institutions, polities and localities as well as between societies and their environments and the dynamics of these phenomena in time and space are visualised and analysed in a qualitative and quantitative as well as comparative perspective. Thereby, the actual complexity of pre-modern societies and the relevance of such research for the analysis of comparable complex interconnections in the contemporary globalised world are rendered visible.

* Correspondence details: Johannes Preiser-Kapeller, Institute for Medieval Research, Division for Byzantine Research, Austrian Academy of Sciences, Wohllebengasse 12-14, 1040 Vienna, Austria, email: Johannes.Preiser-Kapeller@oeaw.ac.at.

The explanatory value of these new methods has been demonstrated with case studies for the Late Byzantine ecclesiastical, political and intellectual elites,[1] processes of religious and ethnic transformations in the Late Medieval Eastern Mediterranean,[2] the diplomatic and political entanglements of the Near East between 300 and 1200 CE,[3] ancient and medieval maritime traffic systems[4] and the complex dynamics of the Late Medieval »World Crisis« in a global and comparative view (see also below).[5] Recently, the *COMMED*-project with regard to its content and methodology was especially augmented by studies on climate and environmental history. *COMMED* focuses not on one single research topic, but provides a methodological and technological framework for various endeavours both within and beyond *IMAFO* (see below).

### Complex systems, networks and mathematics in historical studies

»Systems«, »networks« and »complexity« are terms present in a significant number of current historical and archaeological studies, but in many cases, they are used in a »metaphorical« way or as novel conceptual framework for otherwise traditional narratives.[6] Yet an application of the actual set of concepts and tools provided by the field of complexity, systems and network theory allows for a new understanding, visualisation and analysis of structures and dynamics of past phenomena.[7]

As W. Brian Arthur explains, »complexity is (...) a movement in the sciences that studies how the interacting elements in a system create overall patterns, and how these overall patterns in turn cause the interacting elements to change or adapt«.[8] A »system« consists of interrelated elements, whose interactions at the »micro-level« produce complex changing patterns of behaviour of the entire system on the »macro-level« (»emergence«). For social systems, these patterns stem from the actions and interactions of individuals, families, communities, etc., up to the globalized society of today.[9] These systems show a non-linear behaviour, which means that they answer to certain stimuli (actions of individuals on different scales or external influences and events) or minimal differences in initial conditions not in a linear way (which would mean that the output is proportional to its input); due to the interactions between the parts of the system these stimuli can be reinforced (or weakened) through feedback mechanisms in an unexpected way (»non-linearity«, see also the popularised »butterfly effect«). Furthermore, complex systems are »path-dependent«; their trajectory does not only depend on current conditions, but also on its past dynamics and the structures, constraints and poten-tials emerging from them. Change within complex systems is described as transition between alternative (more or less) stable states or »attractors«. At the same time, complex systems are typically open systems, which are entangled with their environment (consisting of other complex systems both anthropogenic and natural) in equally complex interrelations. The transition from one state to another can emerge because of endogenous interactions (or reactions also to exogenous impacts) which sum up and reinforce (»positive feedback«) or dampen (»negative feedback«) each other until a certain »tipping point« or »bifurcation point« is reached, at which the transition to a new attractor takes place. Marten Scheffer also calls attention to the fact that »systems may gradually become increasingly fragile to the point that even a minor perturbation can trigger a drastic change toward another state«; this depends on »the capacity of a system to absorb disturbance and reorganize while undergoing change so as to still retain essentially the same function, structure, identity, and feedbacks.«, for which Scheffer (as others) uses the term »resilience«.[10] The concept of complex systems is thus deeply »historical« in two respects: it takes into account the system's »history« and at the same time assumes constant dynamic change leading to new system states.[11]

---

1    Preiser-Kapeller, »Our in the Holy Spirit beloved Brothers and Co-Priests«.

2    Preiser-Kapeller, Liquid Frontiers.

3    Preiser-Kapeller, Großkönig, Kaiser und Kalif.

4    Preiser-Kapeller and Daim, *Harbours and Maritime Networks*.

5    Preiser-Kapeller, Networks of Border Zones.

6    Cf. for instance Malkin, *Small Greek World*, and most contributions in: Malkin *et al.*, Greek and Roman Networks. A distinct approach of »system theory« was established by Niklas Luhmann and also applied in another study within the framework of *COMMED*: Preiser-Kapeller, *Luhmann in Byzantium*. For the integration of Luhmann's theories into historical research see also Becker and Reinhardt-Becker, *Systemtheorie*, and Becker, *Geschichte und Systemtheorie*.

7    For an overview cf. also Preiser-Kapeller and Daim, *Harbours and Maritime Networks*.

8    Arthur, *Complexity and the Economy*, 3; Cf. also Mainzer, *Thinking in Complexity*. For a good introduction for historians see Gaddis, *Landscape of History*.

9    On complexity and social theory cf. esp. Castellani and Hafferty, *Sociology and Complexity Science*.

An increasing number of scholars actually refer also to this formal and also mathematical basis, especially in archaeology. Among these approaches, which also have been used within *COMMED*, we find:

Attempts to identify statistical »signatures of complexity« in quantitative data (e.g., distributions of settlements sizes within a region or of wealth among individuals in a larger group, respectively time series, especially of prices), such as unequal distribution patterns (power laws, logarithmic, etc.) or indicators for non-linear dynamics.[12] Furthermore, systems of equations are proposed in order to capture essential factors for these dynamics on the basis of the correspondence between patterns emerging from these models and observed data (a top-down approach).[13]

Experiments to capture the »bottom up«-dynamics of complex systems emerging from the interaction of single elements with the help of agent-based models, acting on the basis of a set of (often relatively simple) rules within a simulated (spatial) environment over several time steps. Again, emerging statistical properties of such models are compared with observed data in order to determine their explanatory value.[14]

Efforts to survey, map and analyse the connections and interactions between various elements (individuals, groups, settlements, polities, but also objects or semantic entities) documented in historical or archaeological evidence with the help of network models in the form of graphs with »nodes« and »ties«, also in their spatial and temporal dynamics. Again, statistical »signatures of complexity« (e. g. patterns of distribution of the number of links among nodes) are identified and models for their emergence in growing or changing networks are proposed.[15]

Many complex systems can be conceptualised as networks of interconnected and interacting elements, and network theory »for many scientists in the community (...) is synonym-

---

ous with the study of complexity«.[16] One central aim of network analysis is the identification of structures of relations which emerge from the sum of interactions and connections between individuals, groups or sites and at the same time influence the scope of actions of everyone entangled in such relations. For this purpose, data on the categories, intensity, frequency and dynamics of interactions and relations between entities of interest is systematically collected in a way which allows for further mathematical analysis. This information is organised in the form of matrices (with rows and columns) and graphs (with nodes [vertices] and edges [links]), which are not only instruments of data collection and visualisation, but also the basis of further mathematical operations (on the basis of matrix algebra and graph theory).[17]

In general, network theory assumes »not only that ties matter, but that they are organized in a significant way, that this or that (node) has an interesting position in terms of its ties.«[18] Differences in the »centrality« of individual nodes can be determined due to the number of their connections or their favourable position between otherwise disconnected groups of nodes. Other methods allow the identification of clusters and cliques as groups of nodes that are more closely intertwined with each other than with the rest of the network, and can represent different factions, for instance. Finally, networks can be analysed in their entirety with regard to the density and »resilience« of the web of relations or the (un)equal distribution of central network positions among nodes.[19] Yet, even in the best cases with thousands of documents, we know for sure that our information is not complete. Written sources provide only a certain part of the spectrum of social relations for a limited group. As for any other historical study, the researcher must be sure that the data basis is sufficient »to demonstrate general structures and developments«[20] – in the case of network analysis, that significant and characteristic patterns can be reconstructed, especially for those types of relationships (kinship, allegiance, economic interaction, etc.), which are essential for the problem at hand. Wolfgang Reinhard, the pioneer of the German historical network research was quite optimistic in this respect: »The selection (of relationship types) in the sources is based on certain rules, for which the values and norms of the historical society were of crucial importance from which the sources originate«.[21] In any case, the systematic collection and presentation of the relations recorded in a stock of sources allows us to discover the big gaps, but also areas of denser evidence, which offer themselves for further structural analysis. On this basis, the actual complexity of social formations of the past and their dynamics become accessible in a new way, as several studies have also demonstrated for the medieval period, especially since the 1990s.[22]

---

16  Johnson, *Simply Complexity*, 13.

17  Wasserman and Faust, *Social Network Analysis*, 92-166; Prell, *Social Network Analysis*, 9-16.

18  Lemercier, Formale Methoden der Netzwerkanalyse, 22. Cf. also Brughmans, Thinking through Networks 623-662, the contributions in Knappett, *Network-Analysis in Archaeology*, and Collar *et al.*, *The Connected Past*, 1-31, for an overview of concepts and tools as well as further bibliography.

19  For an overview of analytical tools see also Preiser-Kapeller, Letters and Network Analysis.

20  Burkhardt, *Hansischer Bergenhandel*. Cf. also Erickson, Social Networks and History; Jullien, Netzwerkanalyse in der Mediävistik.

21  Reinhard, *Freunde und Kreaturen*. For a more pessimistic position cf. Erickson, Social Networks and History, for a well-balanced middle-way see Burkhardt, *Hansischer Bergenhandel*.

22  See for instance: Burkhardt, *Hansischer Bergenhandel*; Gramsch, *Reich als Netzwerk der Fürsten*; Gruber, Wer regiert hier wen?; Habermann, *Verbündete Vasallen*; Mitsiou, Networks of Nicaea; Padgett and Ansell, Robust Action; Preiser-Kapeller, »Our in the Holy Spirit beloved Brothers and Co-Priests«; Sindbæk, Small World of the Vikings; Tackett, *Destruction of the Medieval Chinese Aristocracy*; Vonrufs, *Politische Führungsgruppe Zürichs*. For a critical approach towards these methods cf. Hitzbleck and Hübner, *Grenzen des Netzwerks*. For a review of similar studies in the field of ancient history cf. Nitschke and Rollinger, »Network Analysis is performed.«

*The toolkit applied: mapping medieval conflicts*

Yet, while the term »network« has been used abundantly in historical research in the last years, the actual number of studies taking into account the methodology of network analysis is still relatively limited. The reluctance of historians to adapt tools of network analysis can be also connected with the conceptual and terminological divide between humanities and formal sciences.[23] At the same time, the user-friendliness of software tools tempts others to use them as »black boxes« in order to produce a variety of figures without being aware of the underlying concepts. In order to address these challenges, the expertise in the field of historical network analysis developed within *COMMED* was used by a group of scholars at the *IMAFO* to develop a project which at the same time would provide a comparative approach to medieval history. »*Mapping medieval conflicts: a digital approach towards political dynamics in the pre-modern period*« (*MEDCON*) was selected for funding in September 2014 within the framework of the programme »go!digital« supported by the Austrian Federal Ministry for Science, Research and Economy and the Austrian Academy of Sciences (for the period October 2014 to May 2016).[24] In particular, the aims of *MEDCON* are:[25]

> The adaptation and combination of a set of software tools which facilitates the relational survey of medieval sources and the visualisation and quantitative analysis of social and spatial networks (using an open source database application named »OpenATLAS«, created by Stefan Eichert and developed further together with Katharina Winckler and Alexander Watzinger[26]).
>
> The development of case studies demonstrating a »best practice« of the application and evaluation of tools of network analysis for medieval history.
>
> The creation of an online platform for the exploration of data, methods and results by the wider public.

A generalisable work flow from data input on the basis of medieval sources to the creation, visualisation and analysis of social and spatial network models and their web-based publication and presentation is currently established. In order to demonstrate the explanatory value of a network approach in detail, *MEDCON* focuses on the analysis of political networks and conflict among power elites across medieval Europe with five case studies:

> *Fluctuation between opposing parties in the struggle for the German throne 1198-1208* (Andrea Rzihacek, Renate Spreitzer)[27]
>
> *Coalitions in the war of Emperor Sigismund against Duke Frederick IV of Tyrol* (Günter Katzler)[28]
>
> *Emperor Frederick III Friedrich III. and the League of the Mailberger coalition in 1451/52* (Kornelia Holzner-Tobisch)[29]
>
> *Factions and alliances in the fight of Maximilian I for Burgundy* (Sonja Dünnebeil)[30]
>
> *Political factions in 14th-century Byzantium* (Johannes Preiser-Kapeller).

---

23  Cf. also Hitzbleck and Hübner, *Grenzen des Netzwerks*.

24  http://www.oeaw.ac.at/en/fellowship-funding/promotional-programmes/godigital/ (retrieved 12 September 2015).

25  See also the website of the project: http://oeaw.academia.edu/MappingMedievalConflict (retrieved 25 September 2015).

26  http://www.openatlas.eu/ (retrieved 10 September 2015).

27  Cf. Rzihacek and Spreitzer, *Urkunden Philipps von Schwaben*.

28  For first results see the paper given at the International Medieval Congress 2015, University of Leeds, 6th to 9th July 2015: »Coalitions in the war of Emperor Sigismund against Duke Frederick IV of Tyrol«.

29  Herold and Holzner-Tobisch, *Regesten Kaiser Friedrichs III*.

30  Cf. Dünnebeil, *Orden vom Goldenen Vlies*.

The project is also conceptualised as digital extension of several internationally renowned long-term projects for text edition, diplomacy and prosopography at *IMAFO*.[31]

*MEDCON* uses the relational structuring provided by modern software not simply as instrument for the organisation of data, but as a heuristic tool for the reconstruction and analysis of the relational character of social phenomena of the past.[32] Even if fragmentary tradition does not allow the use of quantitative methods or only to a limited extent, it is worthwhile to take systematically in the focus the social connections between individuals and groups as the context of their actions, especially when it comes to conflicts. Every single actor was embedded in an abundance of relationships which he or she had received by birth (e. g. kinship) or that he or she actively established and maintained (e. g. the membership in a fraternity). These links could be connected with different positions in more or less formalized and institutionalized systems of order (as the patron of followers or as a follower of a higher ranking patron, for instance) and could play an essential role for the identity and overall social position of an individual (e. g. the integration into networks of peers as confirmation of a noble status). In the case of conflict, such networks thus could serve as a resource (support from relatives, friends, allies, patrons), but also limit the room for manoeuvring, due to obligations as a follower of another patron, for instance.[33]

Based on earlier work within the framework of *COMMED*, the additional explanatory value of the approach of *MEDCON* can be demonstrated for the case study on »Political factions in 14th century Byzantium«. As in many parts of Europe, Africa and Asia, also for the Byzantine Empire the »calamitous 14th century« was a time of crisis and conflict. External enemies, climate change associated with natural disasters, and (since the middle of the 14th century) the plague threatened the existence of the empire, while the imperial family and the elite weakened themselves in internal conflicts.[34] The first of these civil wars in the years 1321 to 1328 owed its outbreak to the alienation between the reigning emperor Andronikos II (r. 1282-1328) and his grandson (and formally co-emperor) Andronikos III Palaiologos, who was disinherited. But Andronikos III found his own retinue especially among the younger representatives of the Byzantine aristocracy, who were extremely dissatisfied with the regime of the elder Andronikos who had ruled for almost 40 years with rather limited success. In the spring of 1321, the younger Andronikos and his followers declared war on Andronikos II and demanded the reinstatement of the heir to the throne. The representatives of the elite had to take sides now, and the fracture lines were drawn between generations within individual families.[35]

A telling example is the family of Theodoros Metochites[36], chief minister and a close confidant of the Andronikos II. While Theodoros remained loyal to Andronikos II in the civil

---

war, his sons Nikephoros, Demetrios and Michael[37] established ties with the camp of the younger Andronikos. Furthermore, they also negotiated with their brother-in-law Ioannes Palaiologos, another rebellious scion of the imperial house and husband of Theodoros Metochites' daughter Eirene.[38] Eirene's and Ioannes Palaiologos' daughter Maria had married the Serbian King Stefan Uroš III Decanski[39], who tried to benefit from the internal turmoils of his Byzantine neighbours by backing an attempt of Ioannes to establish himself as ruler of Thessalonike. Thus, the family was separated among three different factions in the years between 1321 and 1328 *(see Fig. 1)*.

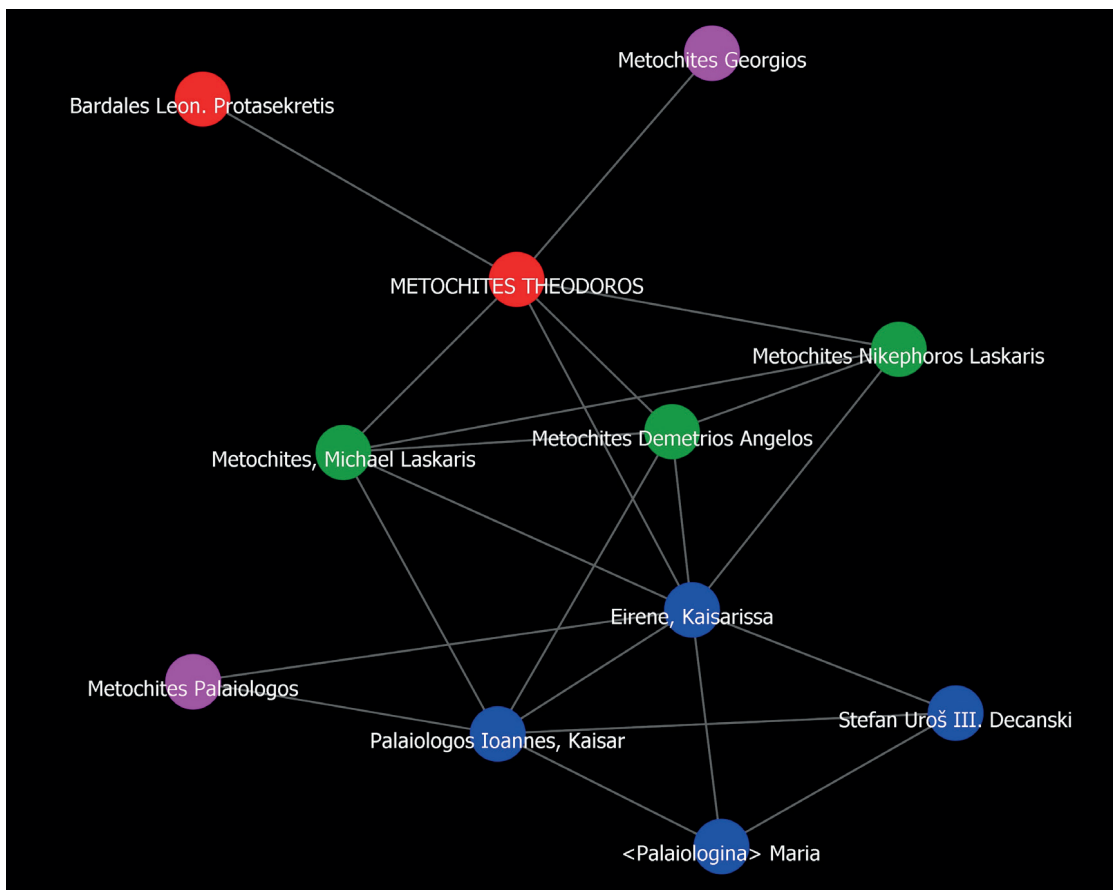

*Fig. 1: The kinship ego-network of Theodoros Metochites in the years 1315-1328 CE; nodes are coloured according to their loyalty in the civil war of the years 1321 to 1328 CE (red: faction of Andronikos II Palaiologos; green: faction of Andronikos III Palaiologos; blue: support of the kaisar Ioannes Palaiologos resp. the Serbian King Stefan Uroš III Decanski; image: Johannes Preiser-Kapeller, 2015, created with the software tool ORA\*)*

Ties of kinship per se thus did not guarantee the cohesion among core groups of the Byzantine elite in that period; still, they provided the most important social bond among the noble clans.[40] As a structural-quantitative analysis of network models for the entire web of kinship among the Byzantine nobility (centred onto the emperor) for several time periods (1282-1292, 1293-1302, 1303-1312, 1313-1321, 1322-1328) during the reign of Andronikos II indicates[41], we can observe underlying general trends towards polarisation in this group even before the outbreak of the civil war in 1321. Our model is not a genealogical one, were every possible kinship tie, which can be reconstructed from the sources, is registered. Beyond the immediate family, where the relevance of kinship is evident (parents and children, siblings, etc.), we only included kinship ties explicitly mentioned as socially relevant and salient in the sources (»my imperial cousin«, »my noble aunt«), thus ties that actually mattered as framework of interactions *(see Fig. 2)*.[42]

---

37 Trapp, *Prosopographisches Lexikon der Palaiologenzeit*, nos. 17980, 17985, 17986.

38 Trapp, *Prosopographisches Lexikon der Palaiologenzeit*, nos. 5972, 21479.

39 Trapp, *Prosopographisches Lexikon der Palaiologenzeit*, nos. 21391, 21181.

40 Cf. Stathakopoulos, Dialectics of Expansion and Retraction (with further literature).

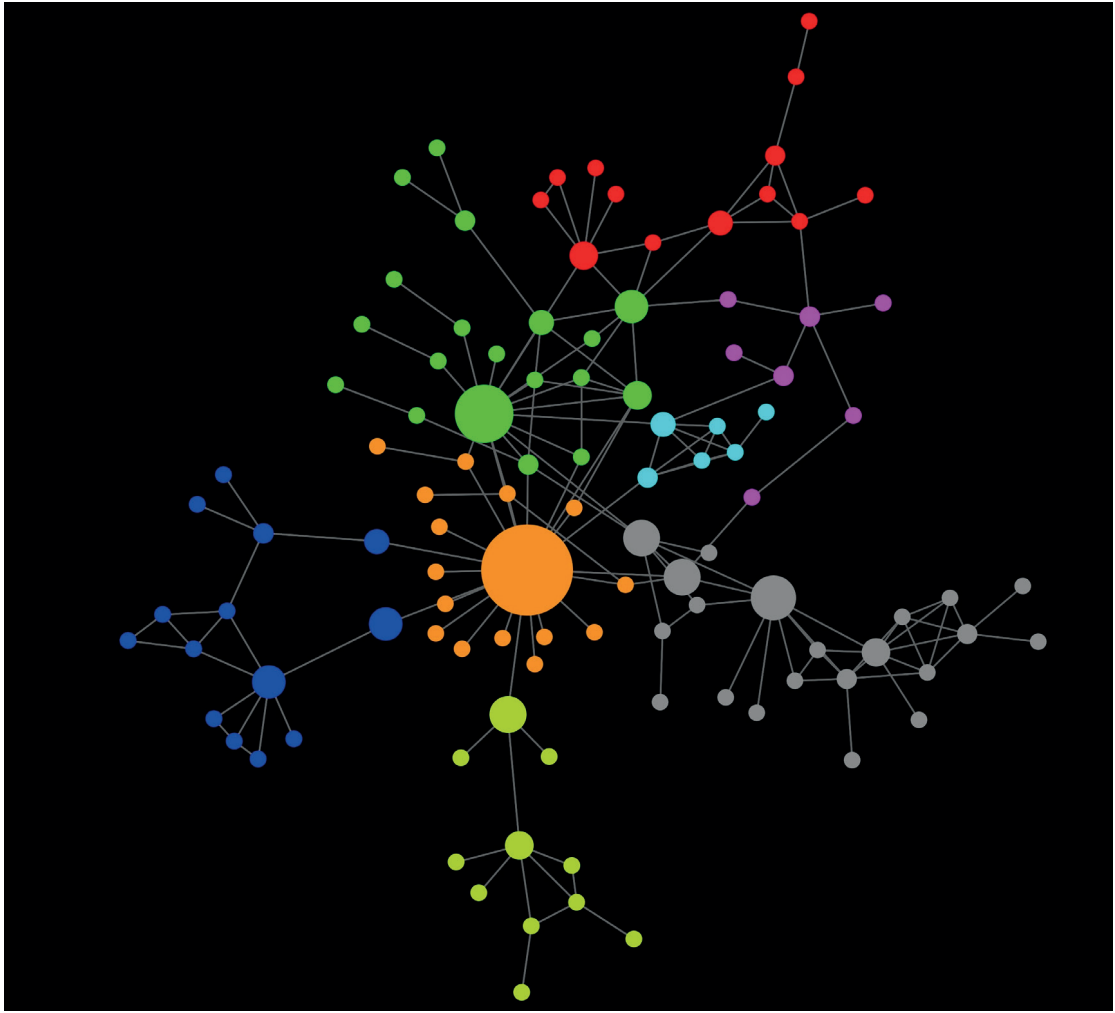

*Fig. 2: Network model of kinship ties between core members of the Byzantine elite (number of nodes = 107) centred on Emperor Andronikos II Palaiologos in the years 1321 to 1328 CE; nodes are coloured according to the cluster identified with the help of the Newman-algorithm (image: Johannes Preiser-Kapeller, 2015, created with the software tool ORA\*)*

41  The network model is based on the data in the »The Vienna Network Model of the Byzantine Elite, 1282-1402«: http://www.academia.edu/8247283/A_new_view_on_a_century_of_Byzantine_history_The_Vienna_Network_Model_of_the_Byzantine_Elite_1282-1402 (retrieved 20 September 2015).

42  Cf. also Vonrufs, *Politische Führungsgruppe Zürichs*.

Across all time periods we observe that the network models of (salient) kinship ties are relatively densely woven; on average, everyone is related to everyone else via »four degrees of separation« (measured in the »average path length«) *(see Fig. 3a and 3b)*. This »macro« network property intensifies on the »micro« level, as is visible in the increasing values for »transitivity«, which indicates the percentage of pairs of links (A to B; B to C) where, when node A is linked to node B, and node B is linked to node C, node C is also linked to node A (»kin of my kin is also my kin«); this process (also called »triadic closure«) increased the »local« cohesion among actors *(see Fig. 3c)*.[43] Equally, we observe a continuous decrease of the values for »degree centralisation« and »betweenness centralisation«, which signal the extent to which ties (»degree«) or the potential for intermediation (»betweenness«) are focused on some especially central actors (such as the emperor); lower centralisation values equal a higher potential for the emergence of other focal points of influence within the network *(see Fig. 3d and 3e)*.[44] (Dis-)assortativity, finally, constitutes a measure to quantify the general amount of structural polarisation, meaning the tendency of nodes to cluster around central actors with a small number of direct connections between these (potentially) »big players«.[45] This measure decreased in the models for the first period of Andronikos' II reign, but increases towards the earlier level until before the outbreak of the civil war *(see Fig. 3f)*. The diremption between the younger and the older Andronikos thus catalysed an already growing trend towards the formation of potentially competing factions within the elite.

Dis-assortativity further increases during the period of civil war 1321-1328 before abating towards the pre-war level during the reign of Andronikos III (r. 1328-1341), who finally was victorious and dethroned his grandfather *(see Fig. 3f)*. Yet, as the values for transitivity and centralisation signal, the underlying dynamics of elite-polarisation did not change during Andronikos' III rule *(see Fig. 3c, 3d and 3e)*; on the contrary, the tendency towards cluster-formation significantly increases during these years *(see Fig. 3g)*.[46] This increasingly delicate balance collapsed when Andronikos III unexpectedly died in June 1341. We »simulated« this event by eliminating the »imperial« node from the network model; the result is a significant increase in the dis-assortativity level and a dramatic decrease of centralisation values *(see Fig. 3f, 3d and 3e)*.

---

43   Nooy *et al.*, *Exploratory Social Network Analysis*, 341-343.

44   Nooy *et al.*, *Exploratory Social Network Analysis*, 143-145, 150-151.

45   Newman, Assortative Mixing. Actually, (degree-based) dis-assortativity is normally measured in negative numbers as counterpart of »assortativity«; here, for the sake of simplification, we measure dis-assortativity in positive numbers.

46   Nooy *et al.*, *Exploratory Social Network Analysis*, 341-343, on the »clustering coefficient«.

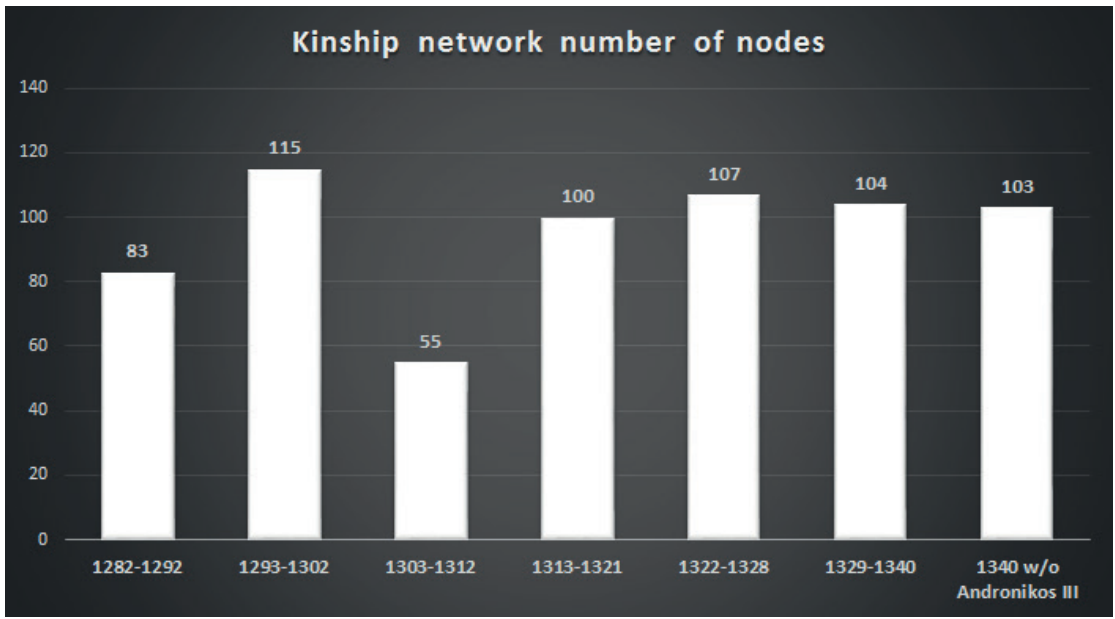

Fig. 3a: Number of nodes in the network models of kinship ties between core members of the Byzantine elite for various time slices between 1282 and 1340 CE (graph: Johannes Preiser-Kapeller, 2015, created with the software tool MS-Excel 2013)

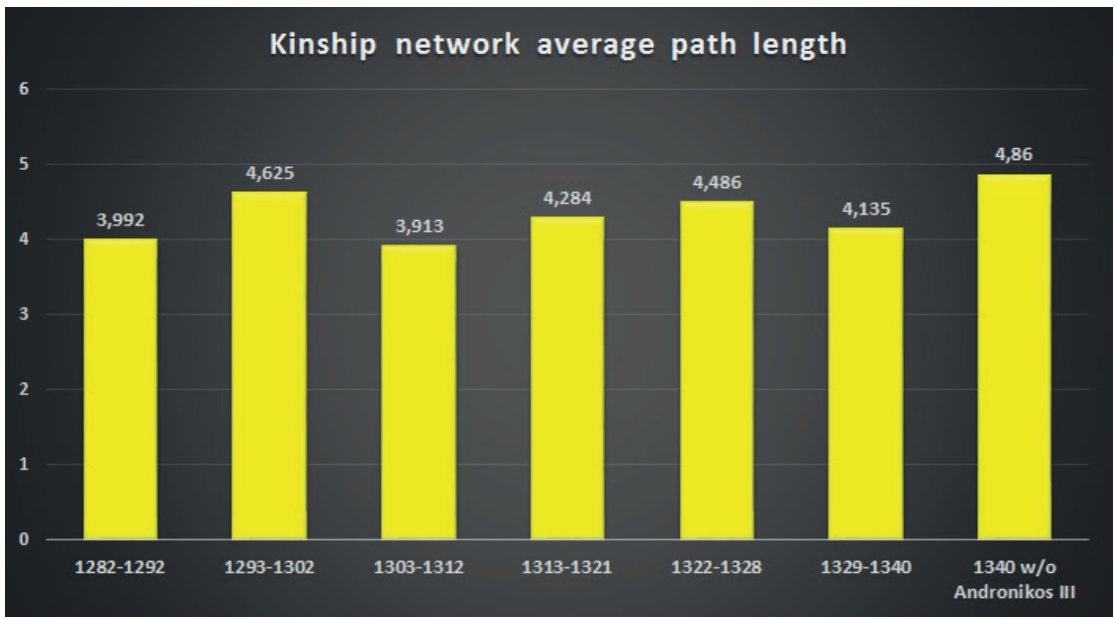

Fig. 3b: Average path lengths between nodes in the network models of kinship ties between core members of the Byzantine elite for various time slices between 1282 and 1340 CE (graph: Johannes Preiser-Kapeller, 2015, created with the software tool MS-Excel 2013)

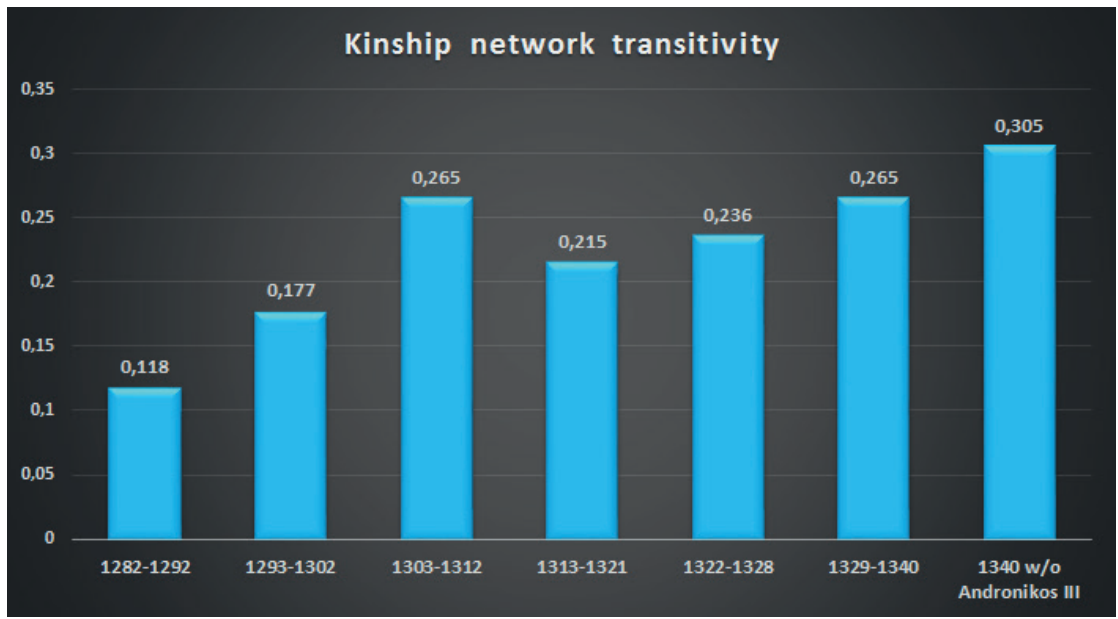

*Fig. 3c: Values of transitivity in the network models of kinship ties between core members of the Byzantine elite for various time slices between 1282 and 1340 CE (graph: Johannes Preiser-Kapeller, 2015, created with the software tool MS-Excel 2013)*

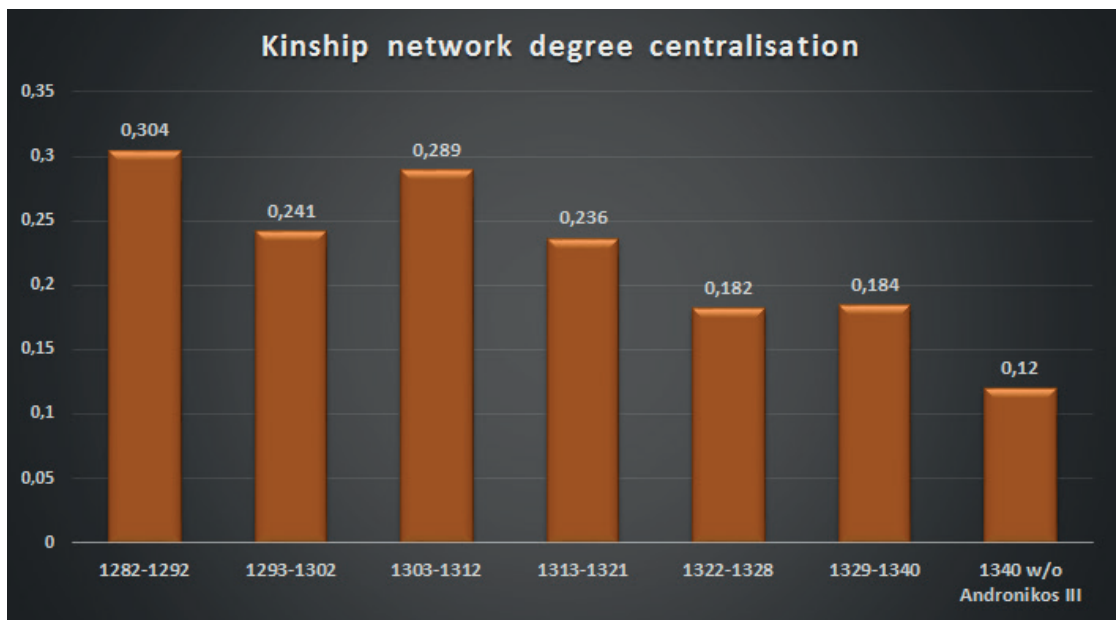

*Fig. 3d: Values of degree centralisation in the network models of kinship ties between core members of the Byzantine elite for various time slices between 1282 and 1340 CE (graph: Johannes Preiser-Kapeller, 2015, created with the software tool MS-Excel 2013)*

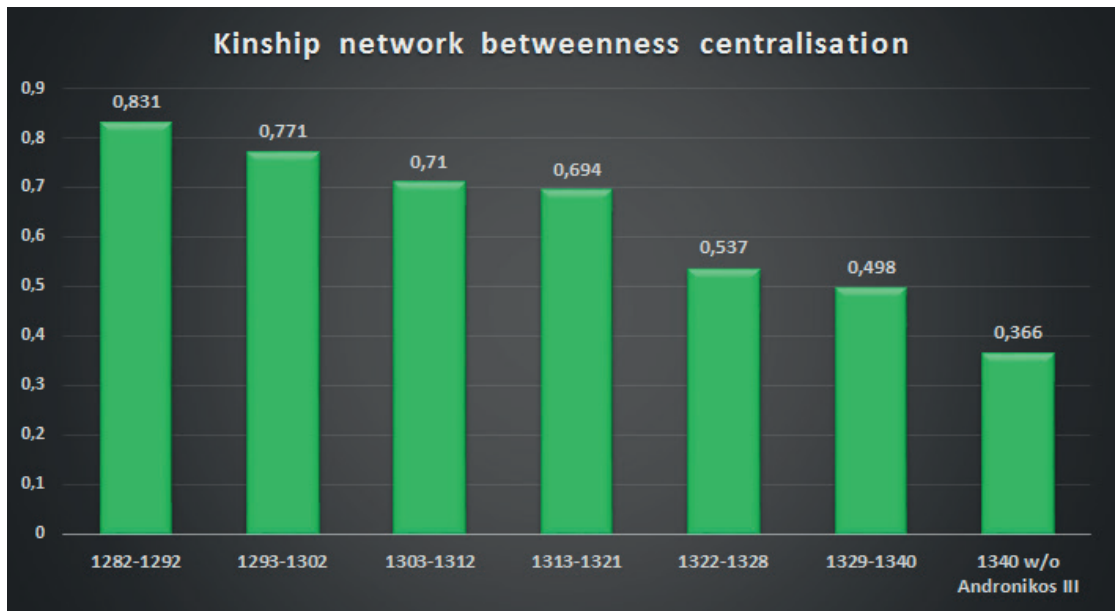

*Fig. 3e: Values of betweenness centralisation in the network models of kinship ties between core members of the Byzantine elite for various time slices between 1282 and 1340 CE (graph: Johannes Preiser-Kapeller, 2015, created with the software tool MS-Excel 2013)*

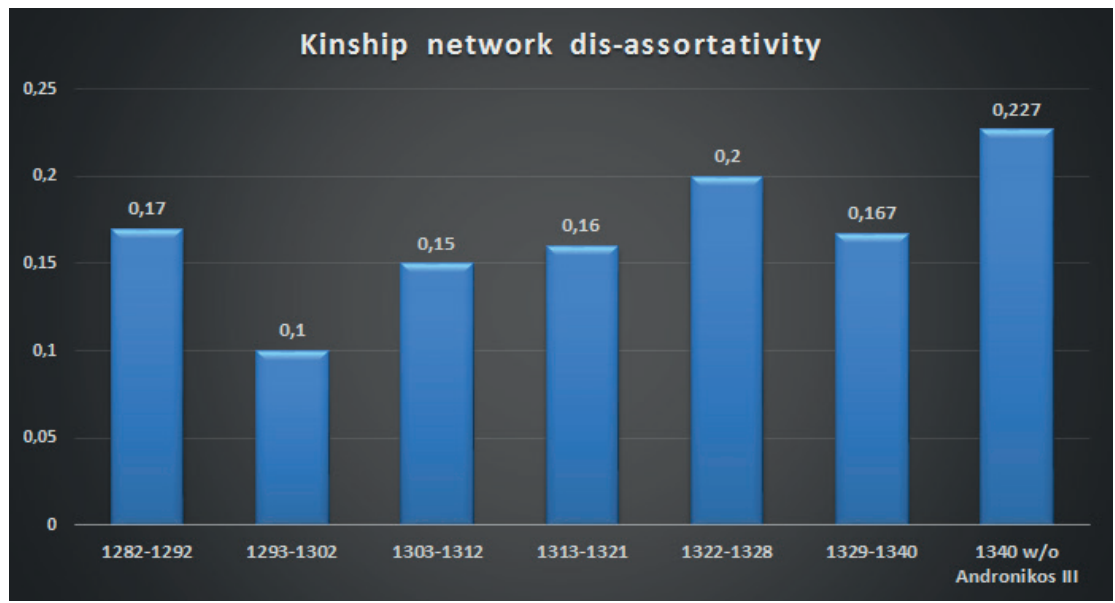

*Fig. 3f: Values of (degree-based) dis-assortativity in the network models of kinship ties between core members of the Byzantine elite for various time slices between 1282 and 1340 CE (graph: Johannes Preiser-Kapeller, 2015, created with the software tool MS-Excel 2013)*

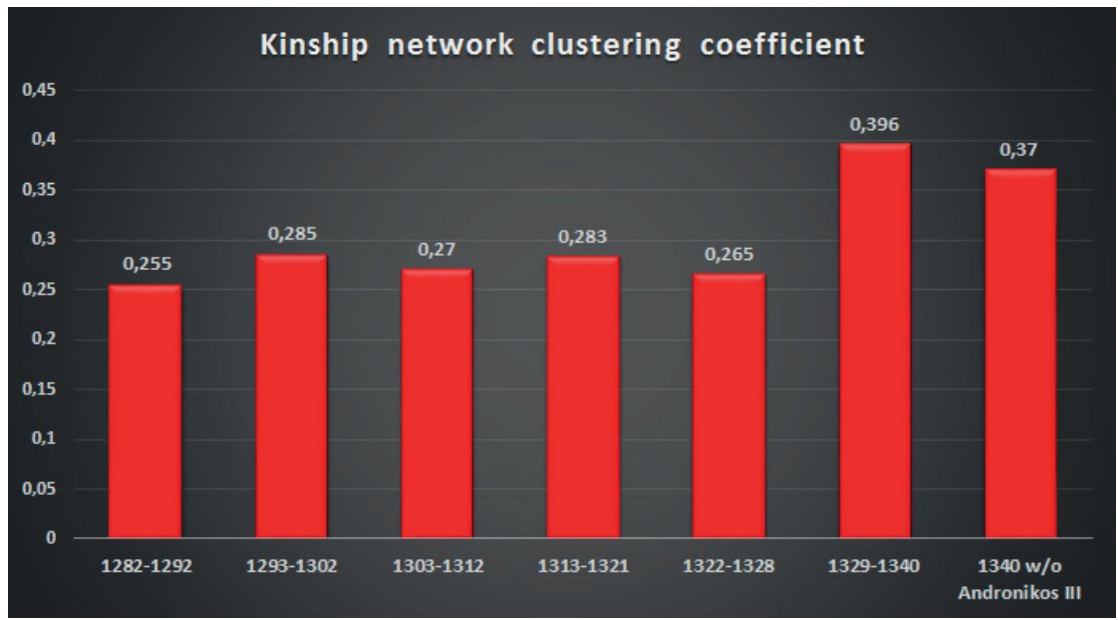

*Fig. 3g: Values of clustering coefficient in the network models of kinship ties between core members of the Byzantine elite for various time slices between 1282 and 1340 CE (graph: Johannes Preiser-Kapeller, 2015, created with the software tool MS-Excel 2013)*

This now salient fragmentation of the elite network formed the structural background for a next round of civil wars in the years of 1341 to 1354, which led to a permanent weakening of the Byzantine Empire.[47] This outburst of intra-elite violence was the most significant one during the 14th century, as a time series of the annually newly added »ties of conflict« between actors in the entire network model for the Byzantine nobility for the time 1282 to 1402 demonstrates *(see Fig. 4)*. In addition, the distribution of frequencies of the number of conflict ties activated in a year tends to follow a power law (characterised by a big number of small-scaled events and a small number of big-scaled events), which is commonly interpreted as a statistical »signature of complexity« and has been observed for other frequency distribution of magnitudes of conflict events *(see Fig. 5)*.[48] Further analysis indicates persistence effects, meaning that the magnitude (in terms of conflict ties) of outbursts of conflict can be correlated with the size of earlier events due to feedback dynamics of polarisation, conflict and (thereby increased or dampened) polarisation within the elite network. The social system of the Late Byzantine elite had a »memory« with regard to conflicts, which influenced the severity of further conflicts.[49]

---

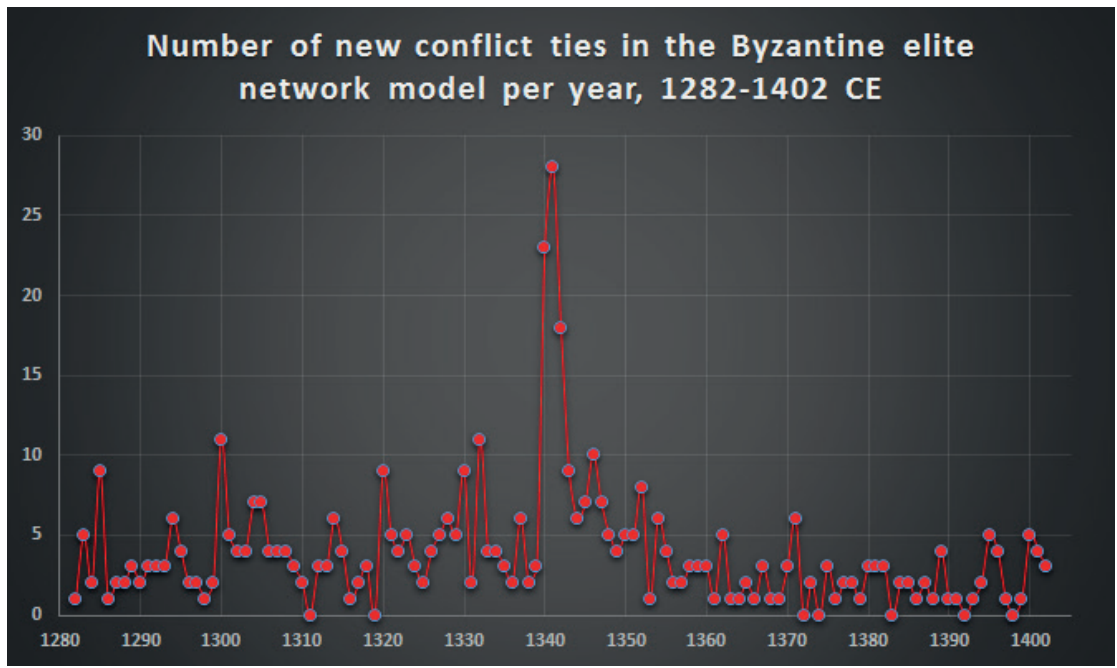

*Fig. 4: Time series of the annually added »ties of conflict« between actors in the entire network model for the Byzantine nobility for the years 1282 to 1402 CE (graph: Johannes Preiser-Kapeller, 2015, created with the software tool MS-Excel 2013)*

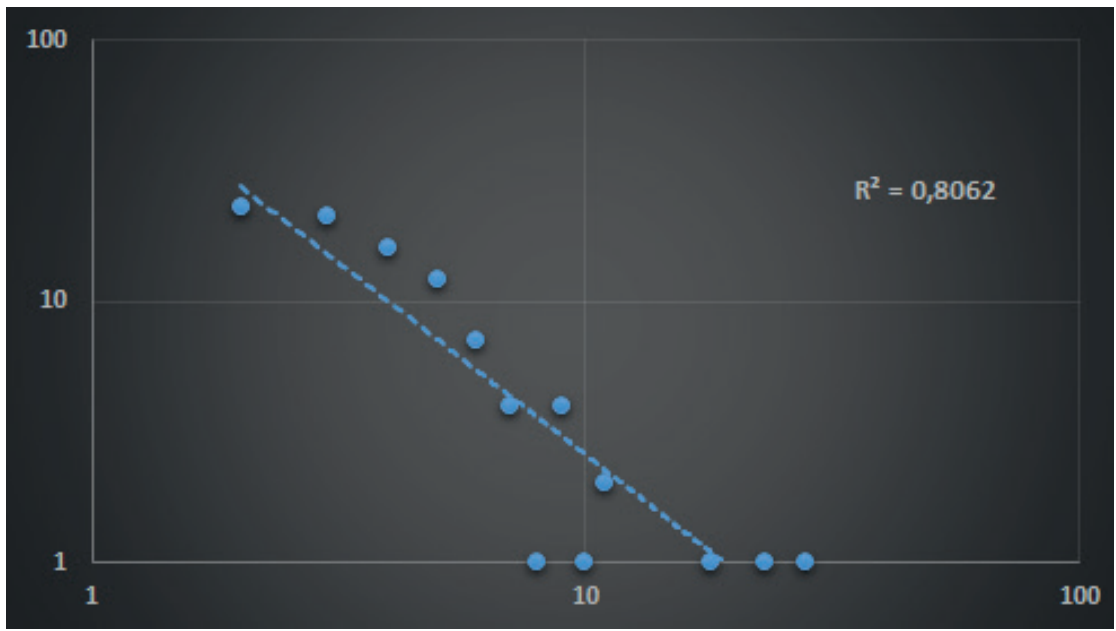

*Fig. 5: Frequency distribution of the number of annually newly added »ties of conflict« between actors in the entire network model for the Byzantine nobility for the years 1282 to 1402 CE (on a double-logarithmic scale; graph: Johannes Preiser-Kapeller, 2015, created with the software tool MS-Excel 2013)*

As *MEDCON* follows a comparative approach, we also ask if the complex dynamics and vulnerability of elite formations were unique to Byzantium. While we are currently creating comparative dynamic network models for the above-mentioned case studies, we inspected the dynamics of internal instability for a sample of five polities from England via Hungary, Byzantium and Egypt to China across Afro-Eurasia. We systematically registered social disturbances such as rebellions, unrest and civil wars and visualised them for the period 1200-1500 CE in the form of »instability indices« (inspired by the studies of Peter Turchin) *(see Fig. 6a, 6b, 6c, 6d and 6e)*.[50] While periods of crises in these five cases very much differ in their chronological distribution and duration, further statistical analysis indicates a general interplay between occurrences of internal instability and of changes of ruler: domestic turmoil not only endangered a ruler's position, a transition on the throne as such in turn increased the probability for further political changes and the outbreak of periods of unrest. This hints at an underlying vulnerability of elite arrangements which became especially salient in cases of »genealogical accidents« such as the premature death of Andronikos III, but contributed to an inherent risk of all cases of regime change *(see Fig. 7)*.[51] On average across all five polities, a change of ruler in one year increased the probability for another change in the following year threefold (an outlier is Mamluk Egypt, where this probability increased only 1.5 times – but here the general risk to encounter a ruler change at any time was exceptionally high with 14 %[52]) *(see Fig. 8a and 8b)*. Furthermore, once a period of instability began, it had the tendency to last; again on average, the probability to encounter a year of unrest after a preceding year of unrest increases six-fold when compared with the transition period from a stable to an unstable year (again, with Egypt as an outlier with a generally higher probability to encounter a year of unrest) *(see Fig. 9a and 9b)*.[53] These statistical properties signal self-energising, complex dynamics (»positive feedbacks«) of political instability in our sample of late medieval polities similar to the case of Byzantium.[54]

---

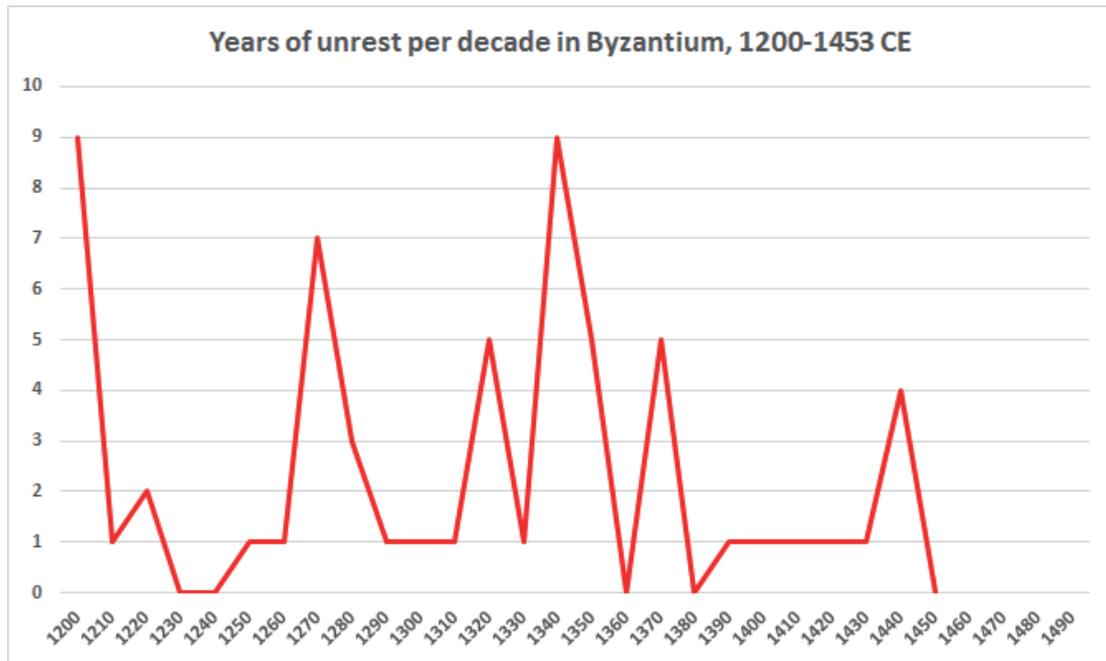

*Fig. 6a: Years of unrest per decade in the Byzantine Empire, 1200-1453 CE (graph: Johannes Preiser-Kapeller, 2015, created with the software tool MS-Excel 2013)*

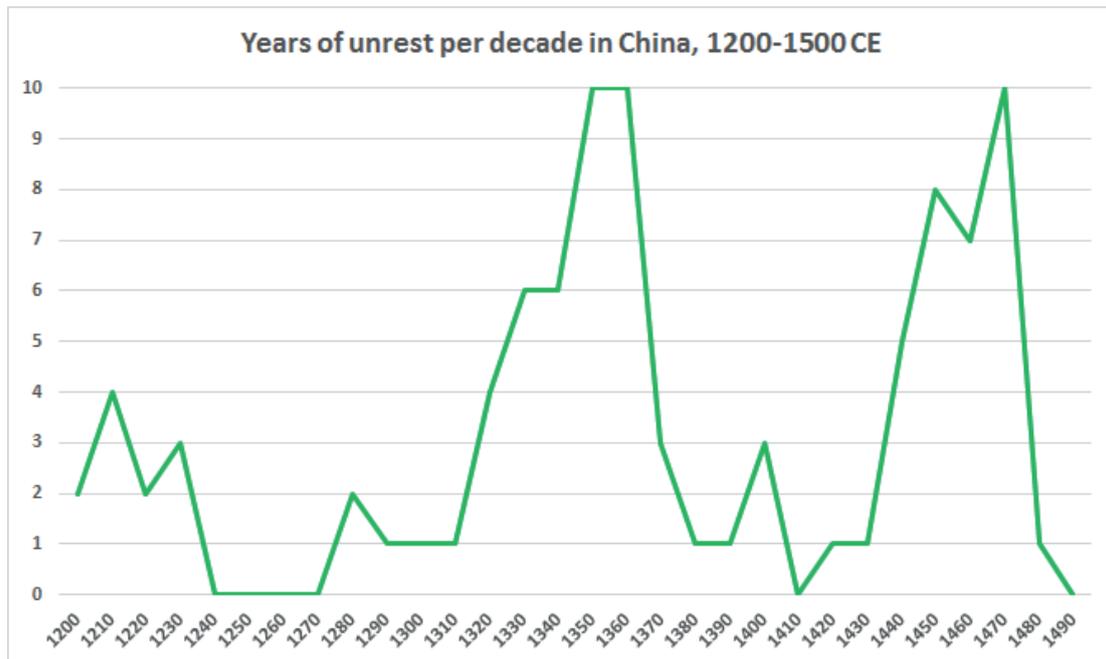

*Fig. 6b: Years of unrest per decade in China, 1200-1500 CE (graph: Johannes Preiser-Kapeller, 2015, created with the software tool MS-Excel 2013)*

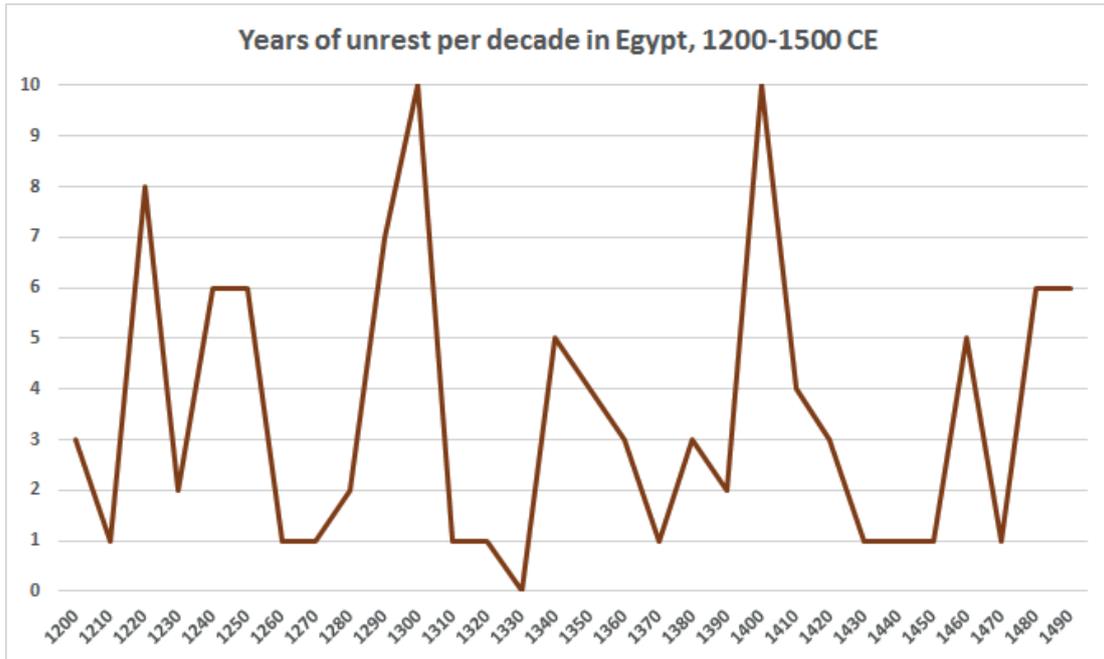

*Fig. 6c: Years of unrest per decade in Egypt, 1200-1500 CE (graph: Johannes Preiser-Kapeller, 2015, created with the software tool MS-Excel 2013)*

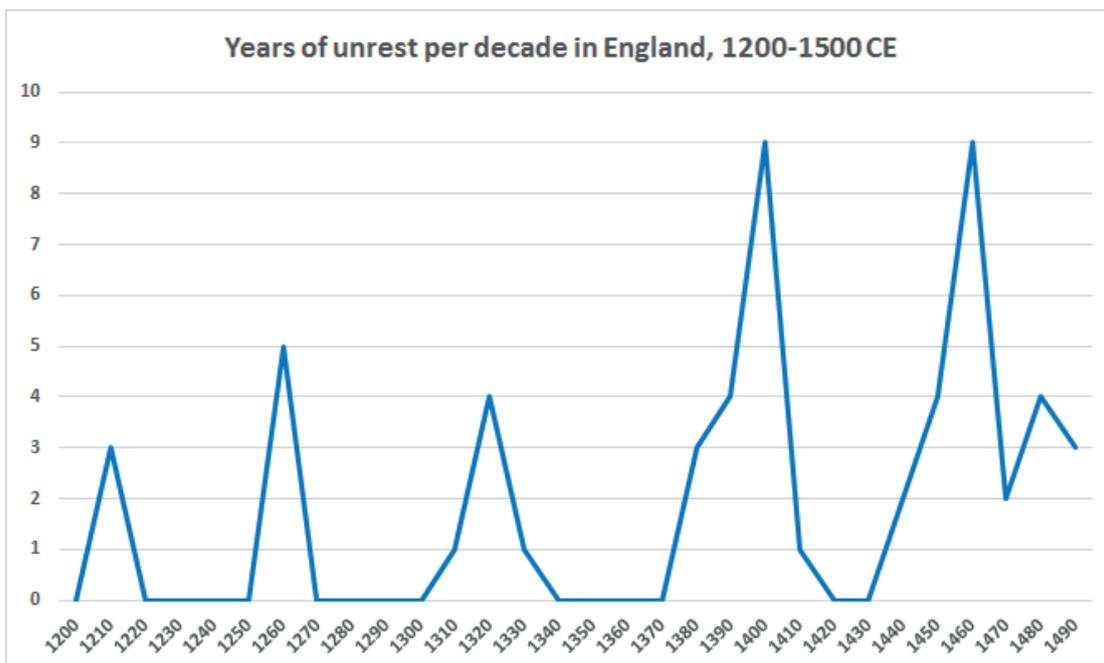

*Fig. 6d: Years of unrest per decade in England, 1200-1500 CE (graph: Johannes Preiser-Kapeller, 2015, created with the software tool MS-Excel 2013)*

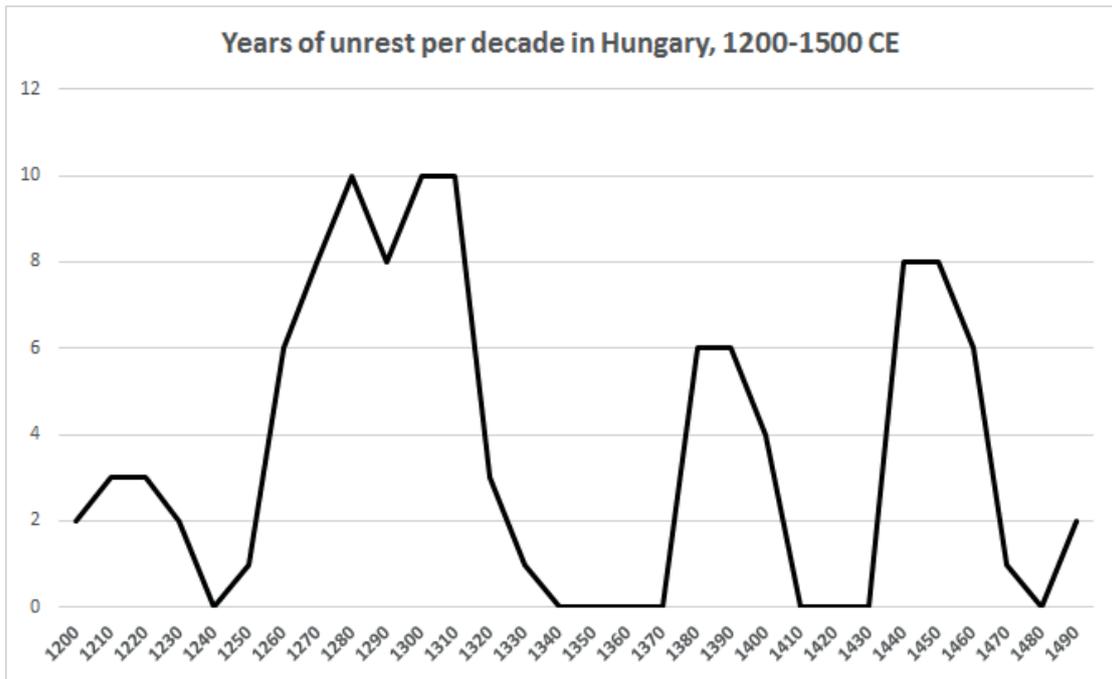

*Fig. 6e: Years of unrest per decade in Hungary, 1200-1500 CE (graph: Johannes Preiser-Kapeller, 2015, created with the software tool MS-Excel 2013)*

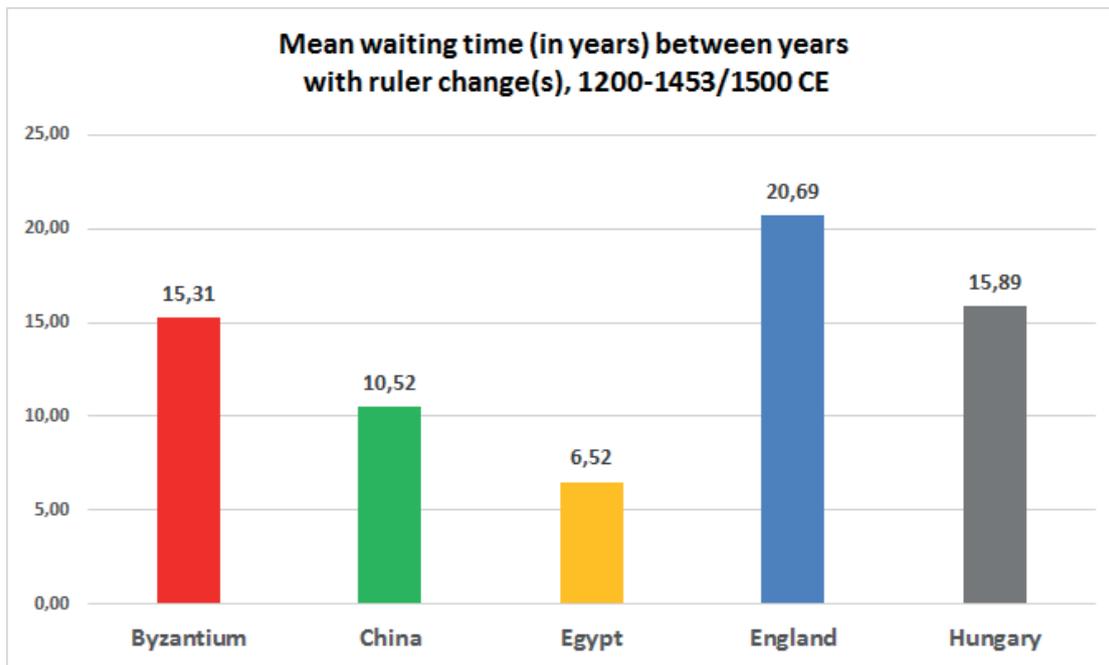

*Fig. 7: Mean waiting time between years with ruler change(s) in Byzantium, China, Egypt, England and Hungary, 1200-1453/1500 CE (graph: Johannes Preiser-Kapeller, 2015, created with the software tool MS-Excel 2013)]*

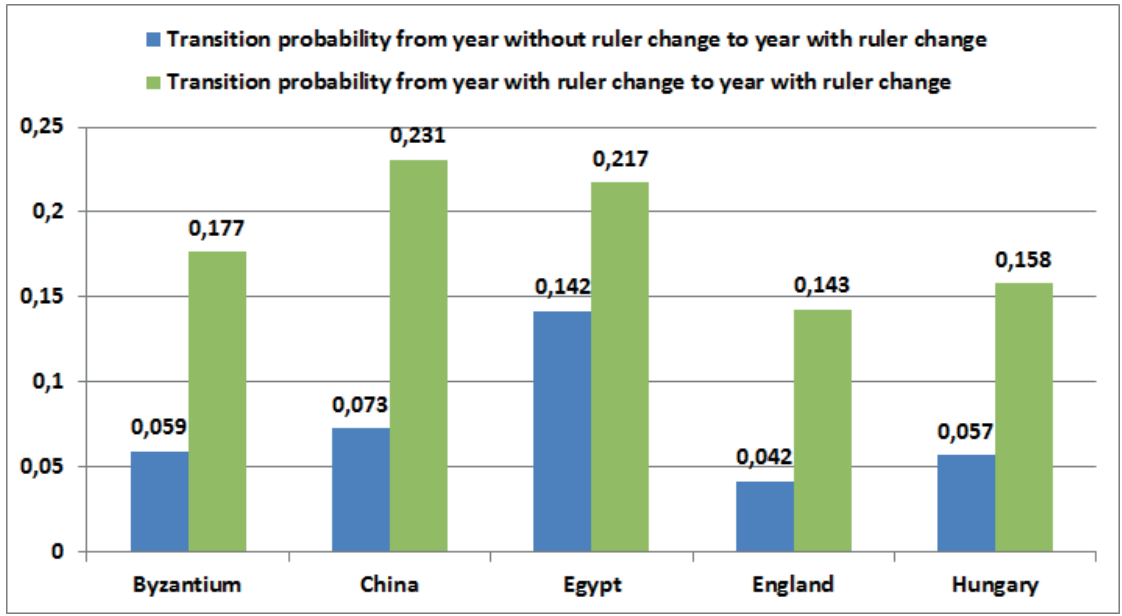

*Fig. 8a: Transition probabilities from a year without ruler change to a year with ruler change respectively from a year with ruler change to another year with ruler change in Byzantium, China, Egypt, England and Hungary, 1200-1453/1500 CE (graph: Johannes Preiser-Kapeller, 2015, created with the software tool MS-Excel 2013)*

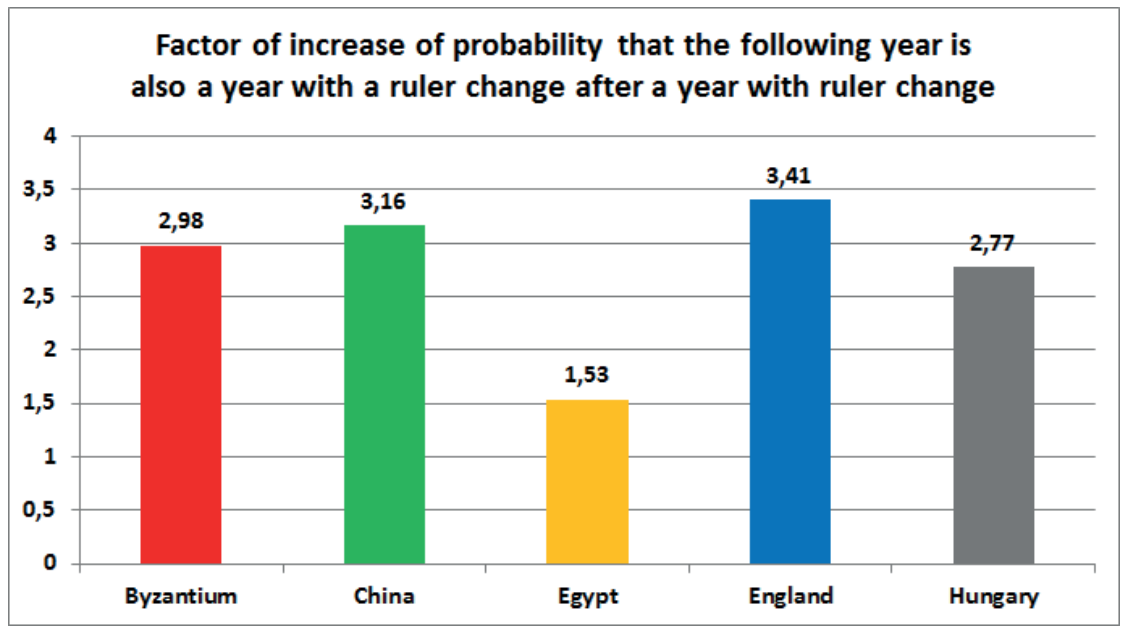

*Fig. 8b: Factors of increase of the probability that the following year is also a year with a ruler change after a year with ruler change in Byzantium, China, Egypt, England and Hungary, 1200-1453/1500 CE (graph: Johannes Preiser-Kapeller, 2015, created with the software tool MS-Excel 2013)*

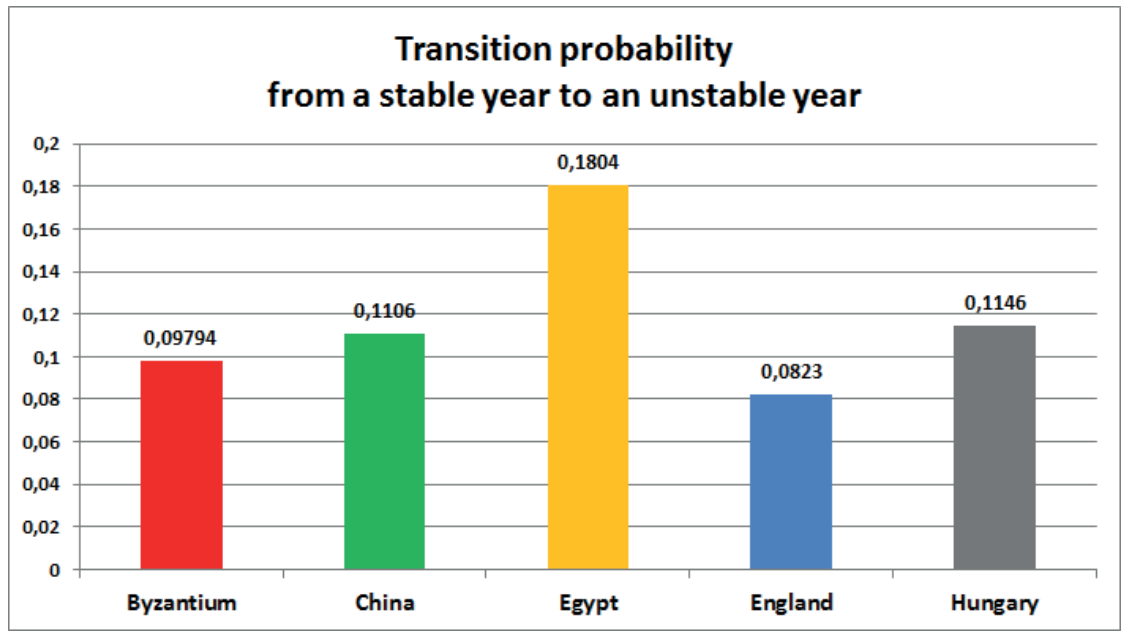

*Fig. 9a: Transition probabilities from a stable year to a year of internal instability in Byzantium, China, Egypt, England and Hungary, 1200-1453/1500 CE (graph: Johannes Preiser-Kapeller, 2015, created with the software tool MS-Excel 2013)*

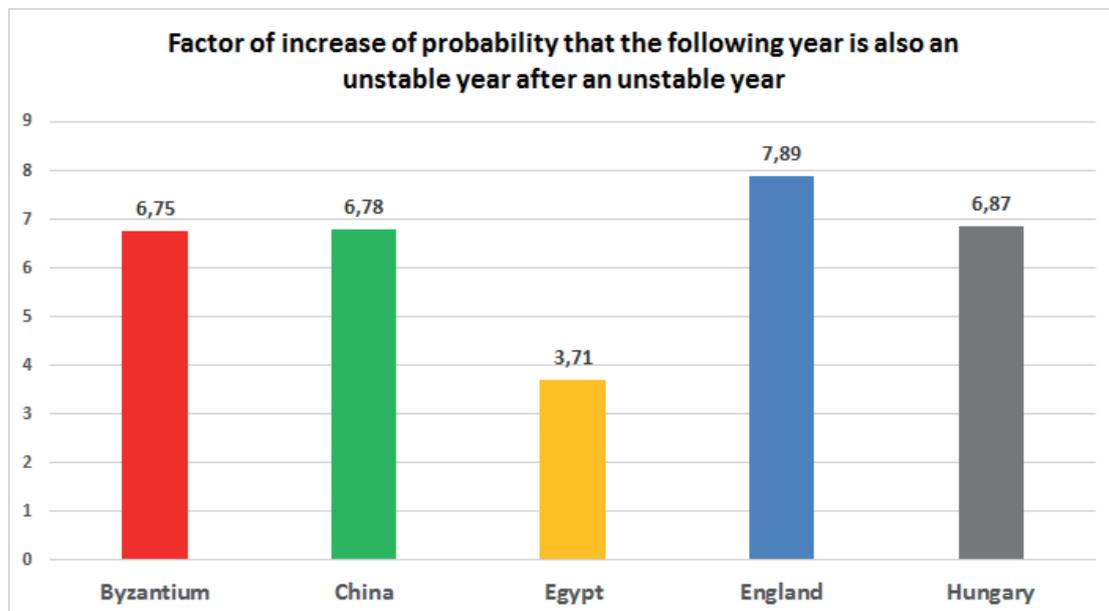

*Fig. 9b: Factors of increase of the probability that the following year is also a year of instability after a year of instability in Byzantium, China, Egypt, England and Hungary, 1200-1453/1500 CE (graph: Johannes Preiser-Kapeller, 2015, created with the software tool MS-Excel 2013)*

Endogenous dynamics in our analysis dwarf the direct (linear) impact of exogenous factors, especially extreme environmental events (droughts, floods, cold snaps) and outbreaks of epidemics, which we registered in a similar way to the instability events *(see Fig. 10a and 10b)*.[55] As indicated above, the late 13th and the 14th century marked the transition from the so-called »Medieval Climate Optimum« (which had positive climatic effects especially on Western Europe, but less so in many other parts of the globe[56]) towards the »Little Ice Age«, accompanied by an increase of the number of extreme events even before the arrival of the »Black Death« and following plague waves from 1345 onwards. This crisis has been identified as a »Schumpeterian wave of destruction« of medieval social arrangements and as a decisive factor for the »transition to modernity«, mostly with regard to Western Europe, whose »rise to global dominance« has been connected to these developments.[57] Therefore, a comparative approach especially beyond Western Europe, integrating polities which »did not make it« to »modernity« (Byzantium, Mamluk Egypt and Hungary, conquered by the Ottomans in 1453, 1517 and 1526 respectively) may allow us to capture the »diversité véritable« without losing track of essential commonalities (the »strange parallels«, as Victor Liebermann has called them in his remarkable study on Southeast Asia in Global Context) of this »world crisis«.[58] Furthermore, a complexity approach provides a more balanced analysis of the interplay between endogenous

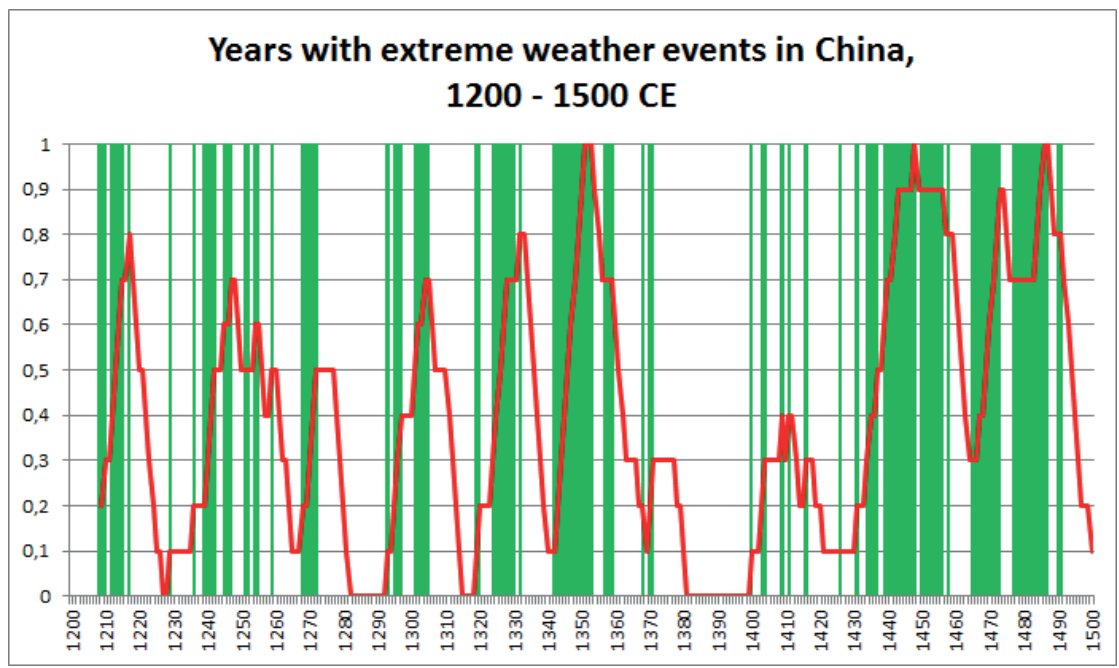

*Fig. 10a: Years with extreme weather events in China, 1200-1500 CE (red: 10 years moving average; graph: Johannes Preiser-Kapeller, 2015, created with the software tool MS-Excel 2013)*

---

social dynamics and exogenous impacts beyond (also recently presented) simplifying scenarios of a linear, maybe even exclusive causation of crises or collapse by forces of climate or epidemiology.[5]

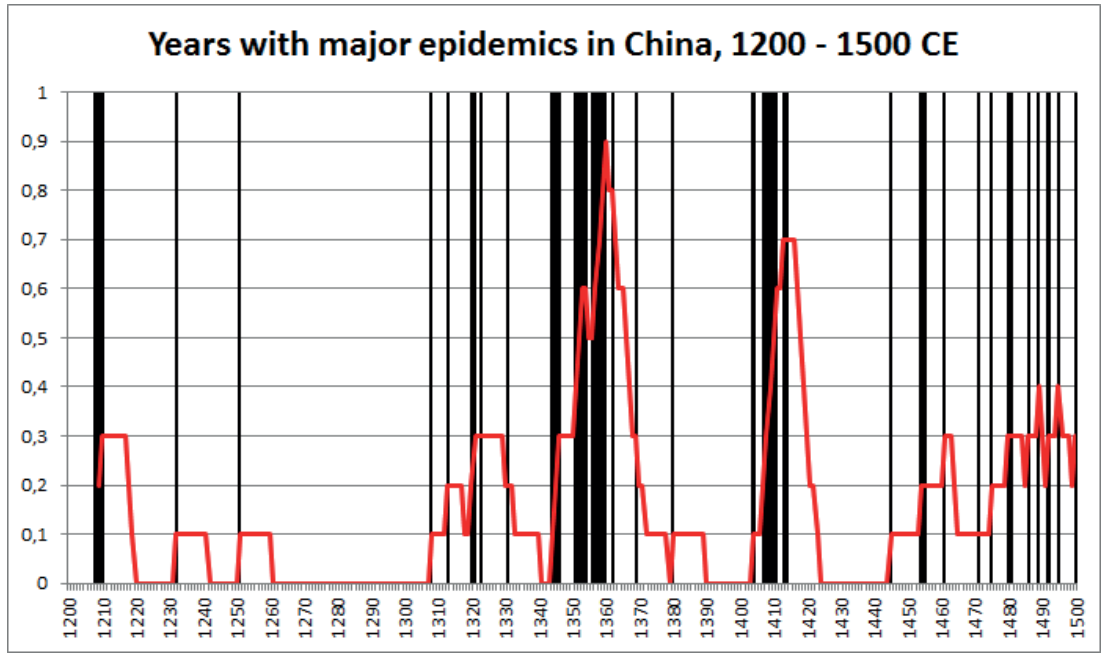

*Fig. 10b: Years with major epidemics in China, 1200-1500 CE (red: 10 years moving average; graph: Johannes Preiser-Kapeller, 2015, created with the software tool MS-Excel 2013)*

### Prospects

After years of methodological development and experiments, the toolkit of COMMED »in action« will be presented in several upcoming publications. Network analysis and complexity theory were combined with climate and environmental history for a book project on the »long 14th century« of the Byzantine Empire (1282-1402) in global perspective, which was accepted by Palgrave Macmillan Publishers under the title »Byzantium's Connected Empire, 1282-1402. A Global History« and shall be published in 2016/2017. This book will be a first monographic synthesis of central methodological and analytical results. In cooperation with Mihailo Popović (OEAW) and Adam Izdebski (University of Cracow) a concept for a first »Companion to the Environmental History of Byzantium« was created, which will be published with Brill in the new series »Companions to the Byzantine World« as a composite work of more than 20 scholars from more than 10 countries.[60] A further book project under the working title »Peaches to Samarkand. Long distance-connectivity, small worlds and socio-cultural dynamics across Afro-Eurasia, 300-800 CE« is in preparation and will focus on the global entanglements between empires and world regions in the transformation period between Antiquity and the Middle Ages, again integrating complexity and network theory with political, socio-economic and environmental history and archaeological evidence.[61]

---

59   Cf. Vries, *Measuring the Impact*; Winiwarter and Knoll, *Umweltgeschichte*; Brooke, *Climate Change*, 391-392, for a recent discussion (with further literature); White, *Climate of Rebellion*. For a detailed analysis of an earlier period of Byzantine history (11th-13th cent.) along these lines cf. Preiser-Kapeller, Collapse of the Eastern Mediterranean?

60   See the outline of the volume: http://www.academia.edu/4098590/A_Companion_to_the_Environmental_History_of_Byzantium_together_with_Adam_Izdebski_and_Mihailo_Popovi%C4%87_eds._ (retrieved 10 September 2015).

As outlined above, on the methodological basis of *COMMED* the project »Mapping medieval conflicts: a digital approach towards political dynamics in the pre-modern period (*MEDCON*)« was awarded funding. Results and further perspectives of network analysis and complexity theory in historical and archaeological research will be discussed at an international conference »Entangled Worlds« in April 2016.[62] Similarly, selected tools are integrated into a case study on historical Southern Armenia (Vaspurakan) for the new project »Digitising Patterns of Power. Peripherical Mountains in the Medieval World« funded within the programme »Digital Humanities: Langzeitprojekte zum kulturellen Erbe 2014« by the Austrian Academy (PI: Mihailo Popović, *IMAFO*).[63]

Within the Division of Byzantine Research, the project is closely connected to the Edition of the Register of the Patriarchate of Constantinople (PRK)[64], the Prosopography of the Palaeologian Period (PLP) and the DFG-funded project »Ports and Landing Places at the Balkan Coasts of the Byzantine Empire (4th-12th Century): Monuments and Technology, Economy and Communication« (together with the Römisch-Germanisches Zentralmuseum in Mainz and the Institute for Byzantine and Modern Greek Studies of the University of Vienna).[65] Beyond the Austrian Academy, the project is cooperating with leading institutions for complexity research in Austria and abroad, such as the »Section for Science of Complex Systems« at the Medical University Vienna (Stefan Thurner), the Department of History, University of Sheffield (Julia Hillner), the Historisches Institut, University of Jena (Robert Gramsch) and the international Evolution Institute with its »Seshat: Global History Databank«, which aims at a macro-comparison of human societies across chronological, spatial and methodological borders.[66] Thereby an exchange of methodological approaches as well as the linking of an increasing amount of data on the political, socio-economic, cultural and environmental dynamics of polities is secured.

*Acknowledgements*
This work was supported by the Austrian Federal Ministry for Science, Research and Economy and the Austrian Academy of Sciences under the Grant scheme »go!digital« (project: »Mapping medieval conflicts: a digital approach towards political dynamics in the pre-modern period«).

---

61  Cf. the working paper Preiser-Kapeller, *Peaches to Samarkand*, with some core arguments and evidence of this project.

62  See: http://www.academia.edu/15158937/Conference_Entangled_Worlds._Network_analysis_and_complexity_theory_in_historical_and_archaeological_research (retrieved 25 September 2015).

63  http://dpp.oeaw.ac.at/ (retrieved 25 September 2015).

64  http://www.oeaw.ac.at/byzanz/prk.htm (retrieved 25 September 2015).

65  http://www.spp-haefen.de/en/projects/byzantine-harbours-on-the-balkan-coasts/ (retrieved 25 September 2015).

66  http://www.complex-systems.meduniwien.ac.at/about/; http://www.sheffield.ac.uk/history/index; http://www.histinst.uni-jena.de/Historisches+Institut.html; https://evolution-institute.org/project/seshat/ (all retrieved 25 September 2015).